\documentclass[]{article}

\def\ben{\begin{equation}}
\def\een{\end{equation}}
\def\bea{\begin{eqnarray}}
\def\eea{\end{eqnarray}}
\input amssym.def
\input amssym.tex
\begin{document}

\hfuzz=100pt
\title{Aspects of Born-Infeld Theory and String/M-Theory}
\author{G. W. Gibbons
\\
D.A.M.T.P.,
\\ Cambridge University,
\\ Wilberforce Road,
\\ Cambridge CB3 0WA,
 \\ U.K.}

\maketitle

\begin{abstract}

\end{abstract}

\section{Introduction}

M/String Theory is the  currently most popular approach to a
unified quantum theory of gravity and the other interactions. We
still lack a complete formulation of the  theory, but there is a
general consensus that whatever finally emerges it will involve in
some way or to some degree of approximation, $p-$ branes, i.e.
$p+1$-dimensional Lorentzian submanifolds $\Sigma$ of a Lorentzian
spacetime manifold $M$. In M-theory one supposes that $M$ is
eleven dimensional. In string theory it is usually taken to be ten
dimensional. Branes may crudely be sub-divided into two types
Heavy and Light.  In the former case one is usually thinking of
many coincident branes whose gravitational field and hence the
ambient spacetime metric  is non-trivial. Semi-classically these
may be studied using supergravity techniques. The other extreme is
to study a single isolated brane moving in flat Minkowski
spacetime as a  solution of the Dirac-Born-Infeld equations of
motion. This will be the approach taken in these lectures. It is
well suited to newcomers  to the subject because, as I will try to
show,  considerable insights into string theory can be gained by
asking some of the simplest physical questions. There is little
need for the full heavy technical machinery of supergravity or
superstring theories. Thus the material is well suited for
presentation at a School. I have deliberately tried to keep things
simple. This runs the risk that experts may feel that I have not
done full justice to the subject or indeed their contributions to
it. If so, I apologize but I repeat my aim was to provide the
beginner with a rapid survey of the subject. I will mainly assume
that the brane is flat. It is fairly straightforward to extend the
present circle of ideas to the case of a curved background. In the
case of Born-Infeld theory the reader is referred to \cite{GK}

The detailed material to be covered is given below.

\tableofcontents

\section{Classical Causality and the Dominant Energy Condition}

A major pre-occuppation of at least  some of the lectures at this school
are  various issues concering causality and locality 
in quantum and classical and in commutative and non-commutative 
field theories. Indeed
they have been part of the motivation of much of the work
reported in in what follows. It therefore seems appropriate to
begin by recapitulating the role of the dominant energy condition,
particularly in the light of some recent papers  either reporting
or theoretically analsying experimental results on the speed of light
and which will be described elsewhere in these proceedings.

\subsection{The Language of Cones}
The appropriate formal language for the discussion is that of convex
cones. Since this seems to be playing an increasingly 
important role in M-theory \cite{GGHT, GWG}
 we shall pause to develop 
it a little. In fact the theory may be developed for general
convex cones, but the most interesting case  is that of homogeneous
self-dual cones. The discussion below will be assuming
that the cones are quadratic but is couched in such a way that it extends
in a straightforwrd way to a  more general setting.

We suppose that an n-dimensional
vector space $X$, ultimately the tangent space $T_xM$
of a spacetime $M$ at some
point $x$, is equipped with a Lorentzian metric $g$. In this section we use
the mainly minus signature $+,-,-,\dots ,-$. 
Picking a time orientation allows us to define the (solid)
 cone $C_ g$ of future
directed causal vectors. In a a time oriented orthonormal basis 
or Lorentz frame 
such that $V=(V^0, {\bf V })$, this consists of vectors
satisfying $V^0 \ge |{\bf V}|$. Conversly a vector $V\in C_g$ iff
$V^0 \ge |{\bf V}|$ in all Lorentz frames. If $W$ is another 
member of of $C_g$  then $g(V,W) \ge 0$ and conversely if
$g(V,W) \ge 0 \quad \forall V \in C_g $ then $W \in C_g$.
One deduces that $C_g$ is a convex cone homogeneous with respect to the
Causal group $\rm {Caus} (n-1,1)$ , i.e. 
the semi-direct product of Lorentz group $SO(n-1,1)$ 
with  ${\Bbb R}_+$ acting as dilations. Thus $C_g= {\Bbb R}_ 
\times SO(n-1,1)/SO(n-1)$.  
The set $Y$ of Lorentzian  metrics $g$ on a fixed vector space $X$
is the homogeneous
space $GL(n , {\Bbb  R} ) /SO(n-1,1)$ and it admits a partial order $<$
(in fact an interesting  type of causal structure)
 which corresponds  to inclusion of 
cones $g < g ^\prime$ iff $C_g \subset C_{g ^\prime}$.
The inclusion need not be strict, i.e. the two cones may touch.  
 
Associated with any convex cone $C \in X$ is the dual cone-cone
or co-cone 
$C^\star$ in the dual  space  $X^\star$. 
This is defined as the set of
covectors $\omega \in X^\star$ such that 
$\omega .V \ge 0 \quad \forall \quad V \in C$.  It is a simple exercise to convince one'self 
that duality reverses
inclusion, $C \subset C^\prime $ iff ${C^\prime } ^\star \subset C^\star$ . 
For a Lorentzian cone $C_g$  the dual cone is given by the inverse metric
$C^\star = G_{g^{-1}}$, 
and  because one may use the metric to set up an isosomorphism between
$X$ and $X^\star$,  the cone is said to be self-dual 
and one does not normally distingush between $C_g$ and ${C_g}^\star$. 
However with more than one metric in the game it is 
essential to make the distinction.

The idea  duality provides a duality between {\it ray} (or particle)
and {\it wave} in all areas of physics. 
The basic observation of De Broglie's 
Ph D Thesis, \cite { DB} may be summarized by saying
was the in Relativity
the  {\sl unique}  Einstein light cone demanded by the 
Equivalence Principle permits an identification
of these two dual concepts and hence leads to Quantum Mechanics.   

\subsection{The Energy Conditions}

Given a metric  may regard the energy momentum tensor
as a billinear form $T_ {\mu \nu}$ or an endomorphism
${T^\mu} _\nu$, according to taste. The various energy conditions
\cite{HE}
depend on the metric. Thus
\subsubsection{The Weak Energy Condition}

One regards  the energy momentum tensor as a quadratic form
and demands that
$T_{\mu \nu} V^\mu V^\nu  \ge 0 \quad \forall \quad V \in C_g$. 
In other words the quadratic form is non-negative on the cone $C_g$.

Because it is equivalent to $T_{00} \ge 0$ in all Lorentz
frames, the weak energy condition is regarded as a fairly mininmal requrement but it is violated
in gauged supergravity theories. This is because they may contain scalar
fields with {\sl negative} potentials. Note that the sign of $T_ {00}$ 
is independent of spacetime signature.

\subsubsection{The Strong Energy Condition}

This is similar and captures the idea that gravity is attractive.
It is used to prove the singularity theorems but
it also implies that any cosmological term must be {\sl negative}
and is inconsistent with inflation. 
It is satisified by all  supergravity models in all dimensions.
Essentially it is incompatible with potentials 
for scalar fields which are {\sl positive}.

One again regards  the energy momentum tensor as a quadratic form
and demands now that 
$(T_{\mu \nu}  -{ 1\over n-2} g _{\mu \nu} T^\sigma _\sigma ){V^\mu V^\nu}
 \ge 0\quad \forall \quad V \in C_g$.   

By contrast

\subsubsection {The Dominant Energy Condition}

is most easilly expressed by regarding the energy momentum tensor
as an endomorphism and demanding that it maps $C_g$ into itself.
That is if $V^\mu$ is causal and future directed 
then so is ${T^\mu}_\nu V^\nu$. Thus $T_{\mu \nu} V^\mu W ^\mu \ge 0
\quad \forall V,W  \in C_g$. An equivalent requirent
is that $T_{00} \ge T_{\mu \nu} \quad \forall \quad \mu \nu $, 
hence the name.

Note that the set $C_{{\rm condition}, g}$ 
of energy momentum tensors satsifying any one of these
conditions with respect to a fixed metric $g$ is itself a convex cone
inside the ${ 1\over 2}n(n+1)$ -dimensional vector space $T$
of 
symmetric tensors.
This accords with one's general prejudice that the state spaces
of physical systems or substances are often convex cones.
The structure of these cones, their boundaries and extreme
points and  mutual dispositions and their dependence on $g$
is an interesting topic
which time does not permit us to pursue in detail here.
We merely remark that one may classify the possible
energy momentum tensors by bringing them to canonical form
( see e.g. \cite{HE} in the case of four dimensions).
Generically one may diagonalize $T_{\mu \nu}$ with respect
to the metric $g_{\mu \nu}$ and we get simple conditions in terms
of the energy density and principal pressures. 
Because the
metric $g_{\mu \nu}$ is not positive definite
there are also some exceptional cases. In this way 
one classifies the orbits of $SO(n-1-,1)$ on the space
of symmetric tensors. Now one may
identify the extreme points, faces etc of the relevant cone.

An important application of the dominant energy condition
is to   the Positive Energy Theorem
of Classical  General Relativity which states that if locally the stress
tensor lies everywhere in the dominant energy cone then the
the ADM energy momentum vector  ${P_{ADM} }^\mu$
of a regular  asymptotically flat spacetime lies in the cone
of future directed causal vectors.

\subsection{{\it Ex nihilo nihil fit}}

We now outline an elegant argument of Hawking \cite{H,HE}
which shows that even if the background metric is time dependent
the dominat energy condition implies causal propagation. If
the metric is time independent this follows form energy conservation
but energy conservation fails if the metric is time dependent
and one might worry that that classically matter might appear "out
of nowhere" that is it might travel at superluminal speeds. The point
of Hawking's argument is that this cannot happen classically.
Of course quantum mechnically things are different, a point    
made strenuously by Zel'dovich \cite{Z1,Z2}. 
Pair-creation processes in external fields
often give the appearance of a-causality because, 
thought of as a tunnelling process, especially using the 
semi-classical or instanton approximation
the particles suddenly materialize at spacelike separations. 
In this context the instanton is often called a bounce.
Think of an electron
positron pair in an external electric field for example.
The instanton is a closed circle in Euclidean spacetime which analytically
continues to a pair of causally disjoint timelike hyperbolae
in Minkowski spacetime \cite{Z1}. If, rather than using the instanton,
one considered a smooth world line  in spacetime with continous 
tangent vector it is clear that in neighbourhood of a creation
event the tangent vector must be spacelike.    

A related point is that the 
quantum mechanical Feynmam propagator
in constrast to the classical retarded or advanced propagator
 has support both inside and {\sl outside} the light cone. As Feynman
has pointed out, this
is because while one may not be able to join two spacelike
points by a smooth timelike curve one may join them by one which
is piecewise smooth and consisting of some past directed and some
future directed intervals. Where the past directed and future directed  
intervals join is the site of a pair-annihilation of or pair-creation event.

Now one might try to describe the pair creation
process using
a regularized expectation value of energy momentum tensor operator,
that is 
\ben
T^{\mu \nu} = \langle {\hat {\bf T}} ^{\mu \nu}   \rangle. 
\een
Zel'dovich and Pitaevsky \cite{Z2} pointed out that the
 dominant energy property
cannot and does not survive the regularization process. 

Here is Hawking's  argument. Let $U$ be a compact region of a spacetime $M$
admitting a time function $t$ whose gradient $\partial_ \mu  t=V_\mu$.
Let the level surfaces of the time function be called $\Sigma _t$
and the part of $U$ earlier than $\Sigma _t$ , i.e. that part
containing events at which the time function
is less than $t$ is called $U_t$.   
The boundary 
$\partial U$ decomposes 
into three components $\partial U_1$ and $\partial U_2$ on which the normal is non- spacelike and time 
function is decreasing or increasing 
along the outward normal ,  
and $\partial U_3$ with  spacelike normal.
We note that
\ben
{J^\mu}  _{;\mu}= T^{\mu \nu} V_{\mu ;\nu}, \label{divvy}
\een
where, by the dominant energy condition,  $ J^\mu = T^\mu _\nu V^\nu$  is 
a future directed timelike vector field. By 
the dominant energy condition and the compactness of
of $U$ that there exists a positive constant $P$ such that
\ben
T^{\mu \nu} V_{\mu ;\nu} \le P T^{\mu \nu} V_\mu V_\nu.
\een
Let 
\ben
E(t) = \int _{\Sigma_t} J_\mu  \Sigma  ^\mu.
\een
Clearly $E(t) \ge 0$ and $E(t)=0$ implies that $T^{\mu \nu}$
vanishes on $\Sigma_t$. 

Integration of (\ref{divvy}) over $U(t)$ 
gives
\ben
E(t) \le -\int _ {U(t) \cap \partial U_1 } J_\mu d \Sigma ^\mu +
\int _{U(t) \cap \partial U_3} J_\mu d \Sigma ^\mu   + P \int ^t dt ^\prime 
E(t^\prime ).
\een

We have used the fact that by the dominant energy condition
\ben
\int _ {U(t) \cap \partial U_2 } J_\mu d \Sigma ^\mu  \ge 0.
\een

Now suppose that $T^{\mu \nu}$ vanishes on $\partial U_3$,
the timelike
component  of the  boundary $\partial U$ and so nothing flows into the 
region $U$. We deduce that
\ben
{dE \over dt} \le P E(t). 
\een
Integration of this simple this simple differential identity
implies that
\ben
E(t) \le  E(t^\prime) \exp{ P(t-t^\prime )}.
\een
and hence that if $E(t)$ vanishes at some time $t^\prime$
, then  it must vanish for all times.
This despite the fact that we have allowed for the possibility of a time
dependent metric doing work on the matter. One cannot get somethig from
nothing. To get a statement about causality we apply this result
to the case when $U=D^+(S)$ the future Cauchy
 development of some set $S$. This is the set of all points $p$
such that every 
past directed causal curve through $p$ intersects $S$.
If $T^{\mu \nu}$ vanishes on $S$ then it vanishes everywhere in $S$. 
  
The results just given, and obvious generalizations
show clearly that according to Maxwell's equations,
electromagnetic waves can never, in
the sense defined above, 
travel faster than light.

\section{Open Strings and D-branes}

Branes may be incorporated in string theory if one contemplates
opens strings whose ends are constrained
(by Dirichlet boundary conditions) to lie on a
$(p+1)$-dimensional submanifold $\Sigma _{p+1}$.
Now open strings can couple minimally
to vector $A_\mu$  at the ends of the strings.
In the Polyakov approach one has an action of the form
\ben
-{ 1\over 2} \int_{\Sigma _1}
 d^ 2\sigma (G_{ab} + B_{ab}) \partial y^a  \partial y^b
+ \int _{\partial \Sigma _1} A_a dy^a, \een where the embedding of
the string world sheet $\Sigma _1 \rightarrow M$ is given by $y^a
=y^a (\sigma^A)$, $A=1,2$ and $a=1,2, \dots , n={\rm dim} M$, and
$G_{ab}$ and $B_{ab}$ are the spacetime metric and Neveu-Schwarz
two-form respectively.

One obtains an effective action for a D-brane
if one "integrates out" all possible  string motions
subject to the Dirichlet boundary condition.

The resulting action depends on the position of the D-brane and
the pullback to the D-brane of the metric and Neveu-Schwarz
two-form. It also contains the vector field $A_\mu$.

\section{Dirac-Born-Infeld Actions}

This is governs the embedding $y: \Sigma _{p+1} \rightarrow  M$
given in local coordinates by $y^a =y^a( x^\mu)$, where
$a=1,2,\dots, n= {\rm dim} M$ and $ \mu =0,1,2, \dots, p$. It is
\ben - T_p \int dx^{p+1} \sqrt {\det (g_{\mu \nu} + (2 \pi \alpha
^\prime ) F_{\mu \nu} + B_{\mu \nu} )}, \een where \ben g_{\mu
\nu} = \eta_{ab} \partial _\mu y^a \partial _\nu y^b, \een and
\ben B_{\mu \nu} = B_{ab} \partial _\mu y^a \partial _\nu y^b,
\een are the pull-backs of the metric $\eta_{ab}$ and
Neveu-Schwarz two-form $B_{ab}$ to the world volume $\Sigma_{p+1}$
of the $p$-brane.

The the world-volume field $F_{\mu \nu}$ is
given by
\ben
F_{\mu \nu} = \partial _\mu A_\nu - \partial _\nu A _\mu.
\een

One often defines
\ben
{\cal F}_{\mu \nu} = F_{\mu \nu} + B_{\mu \nu}.
\een
This is invariant under a Neveu-Schwarz gauge transformation
$B \rightarrow B-dC$ where $C$ is a one-form, if we transform
$F \rightarrow F+ dC$. One may check that this is consistent
with the behaviour of the open string metric.

\subsection{Monge Gauge}

To proceed we fix some of the gauge-invariance associated with
world sheet diffeomorphisms of the coordinates $X^\mu$ by using
what is usually, and misleadingly  called static gauge (since it
applies in non-static situations) and which is more accurately and
with more justice called Monge gauge. In effect we project onto a
$p+1$ plane by setting $y^a= x^\mu, y^i$ and use the $n-p-1$
height functions $y^i$, $i=1,2,\dots ,n-p-1$ as scalar fields on
the world volume. In the theory of minimal surfaces this is called
a non-parametric representation. For Monge's work see
\cite{Monge}. Of course there may not be a global Monge gauge, and
we shall encounter this situation later.

The determinant then becomes (we use units in which $2\pi \alpha ^\prime=1$),

\ben
\det( \eta_{\mu \nu} + \partial_ \mu y^i
\partial _\nu y^i + F_{\mu \nu} ).
\een
It is evidently consistent to set the scalars to zero $y^i=0$ and we then
obtain the Lagrangian of Born and Infeld which is a
special form of Non-Linear Electrodynamics.

\subsection{Dimensional Reduction}

The previous section result has a sort of converse. We could start
with a pure Born-Infeld action in $n$ flat  dimensions and
dimensionally reduce to $p+1$ dimensions. We begin with \ben -\int
d^nx \sqrt{- \det ( \eta_{ab} + F_{ab} )}. \een We make the ansatz
$A_a= (A_\mu(x^\lambda), y^i(x^\lambda) )$ and obtain the
Monge-gauge-fixed Dirac-Born-Infeld  action \ben -\int d^{p+1}
\sqrt{- \det ( \eta_{\mu \nu} + F_{\mu \nu} + \partial _\mu y^i
\partial _\nu y^i )}.
\een
Thus all solutions of the Dirac Born-Infeld action
are solutions of the Born-Infeld action. Interestingly
in the case $p=1$ we get a string action from the pure Born-Infeld action.

\section{Non-Linear Electrodynamics}

There are advantages in viewing  the theory in this context.
An excellent account of the theory is given in \cite{B1}.
The general
theory in four-spacetime dimensions (p=3) has equations
\ben
{\rm curl}  {\bf E} = -{\partial {\bf B} \over \partial t}: \qquad {\rm div}
{\bf  B}=0  \label{A}
\een

\ben
{\rm curl } {\bf H} = {\partial {\bf D} \over \partial t}: \qquad {\rm div}
{\bf  D}=0 \label{B}.
\een
\subsection{Constitutive Relations}
To close the system one needs constitutive relations
${\bf H} = {\bf H} ({\bf E}, {\bf B} )$ and ${\bf D} = {\bf D} ({\bf E}, {\bf B} )$
which, if one has a  Lagrangian
$
L=L( {\bf E}, {\bf B} )$, take the form
\ben
{\bf H} = -{\partial L  \over \partial
 {\bf B}} \qquad
{\bf D} = {\partial L \over \partial {\bf E}}.
\een
Because
\ben
{\bf D} = {\partial L \over \partial {\dot {\bf A} } },
\een
$\bf D $ is the canonical momentum density. Note also that
the conserved electric charge is given by the flux of $\bf D$
and {\sl not} as is often assumed, the flux of $\bf E$.

In what follows we shall denote by $K_{\mu \nu}$
the Amp\`ere 2-form with components $({\bf D},{\bf H} )$
and refer to $F_{\mu \nu}$ as the Faraday 2-form.
Thus the equations of motion without sources are
\ben
dF=0 \qquad d\star K=0.
\een

\subsection{Lorentz-Invariance}
The symmetry of the energy momentum tensor $T_{0i}=T_{i0}$ and
hence the uniqueness of the Poynting vector requires that the
latter be given by \ben {\bf E} \times {\bf H}= {\bf D} \times
{\bf B}. \een This will follow if $L$ is constructed from the two
Lorentz invariants \ben x={1 \over 2} ({\bf B}^2 -{\bf E}^2) \een

\ben
y= {\bf E} \cdot {\bf B}.
\een

\subsection{Duality Invariance}

The constitutive relations will permit
the obvious rotation needed to
rotate the two sets of equations (\ref{A}, \ref{B}) into themselves
\ben
{\bf E} + i {\bf H} \rightarrow e^{i \theta} ({\bf E} + i {\bf H} ),
\een
\ben
{\bf D} + i {\bf B} \rightarrow e^{i\theta}  ({\bf D} +i {\bf B} ),
\een
with $\theta$ constant
if
\ben
{\bf E} \cdot {\bf B}= {\bf D} \cdot {\bf H}.
\een

Note that what we are encountering here is a {\sl  non-linear form
of the familiar linear Hodge duality} This gives  a constraint on
possible theories. For example if the Lagrangian depends
arbitrarily on the invariants $x$ and $y$ it gives rise to a
Lorentz-invariant theory. Imposing duality invariance reduces this
freedom to that of a function of a single variable.
For more details on duality invariance see \cite{GR1, GR2} and \cite{B1} which was not known to the authors
of \cite{GR1,GR2} when they were written.

\subsection{Hamiltonian density}

One has
\ben
{\cal H} = T_{00}= {\bf E} \cdot {\bf D} -L.
\een
one may think of ${\cal H} =
 {\cal H} ({\bf B}, {\bf D} )$ as
the Legendre transform of the Lagrangian  and is thus expressed in
terms of the canonical variables  $\bf B$ and $\bf D$ whose
Poisson Brackets are \ben \{ B_i( {\bf x}) , D_j ({\bf y} )  \} =
-\epsilon_{ijk} \partial _k \delta({\bf x}-{\bf y} ). \een

\subsection{Born-Infeld}
We have \ben L= 1-\sqrt {1-{\bf E} ^2 + {\bf B} ^2 - ({\bf E}
\cdot {\bf B})^2 } \een and \ben {\cal H} = \sqrt{ 1+{\bf B} ^2 +
{\bf D} ^2 + ( {\bf B} \times {\bf D} )^2   }-1. \een A constant
has been added to make the zero field have zero energy. This is
not strictly necessary in the theory of banes since the notion of
world volume energy is not well defined because there are no
privileged coordinates on the brane. However  it is convenient
when making comparisons with standard flat space field theory. To
do so we must however use Monge gauge.

Lorentz and Duality invariance are clear. Before the advent of
String/M-theory the latter  was rather mysterious.  Nowadays it
may be thought of as a manifestation of S-duality. In this way we
see how Born-Infeld theory considered {\it sui generis} has
important lessons for M/String theory. Conversely M/String theory
throws light on Born-Infeld theory. We shall see more examples of
this mutually symbiotic behaviour later.

\section{The Maximal Electric Field Strength}

If ${\bf B}=0$, the Born-Infeld Lagrangian is
\ben
L=1 -\sqrt {1- {\bf E}^2  }.
\een
If we use a gauge in which $A_0=0$, we have
\ben
L=1 -\sqrt {1 -{\dot {\bf A}} ^2  }.
\een
The analogy with special relativity is clear. There will be an
upper bound to the electric field strength.
The special relativistic analogy  may also be understood from
the point of T-duality.

In string theory the existence of a maximal electric field
strength  may be understood dynamically as follows. A stretched
open string of length $L$ has, in our units, elastic energy $L$.
If it has charges $+1$ at one end and $-1$ at the other it will,
in an electric field have energy $-EL$. This if $E>1$ one may gain
energy from the background electric field by creating open
strings. This an electric  field with strength greater than $1$
will quickly breakdown and the electric field will be reduced to a
value less than one.

Note that if one restores dimensions and units the critical field
strength $E_c$  is given by \ben E_c = { 1\over 2 \pi \alpha
^\prime}. \een In the zero slope limit $\alpha ^\prime \rightarrow
0$ there is no upper bound and in the strong coupling limit
$\alpha ^\prime \rightarrow \infty$ the critical field goes down
to zero. Later we will investigate the behaviour of the theory in
this limit.

\section{BIons}

The maximal electric field was originally invoked to ensure the
existence of a classical solution representing a charged object
with finite total energy \ben \int_{{\Bbb E}^3} d^3x T_{00} <
\infty \een This can be achieved by setting \ben {\bf D} = { q
\over r^2} {\hat {\bf r}}. \een Because \ben {\bf E}= { {\bf D}
\over \sqrt{1+ {\bf D} ^2 } } \een the electric field achieves its
maximal value at the centre. Note that \ben {\bf D}= { {\bf D}
\over \sqrt{1-{\bf D} ^2 } } \een the electric induction ${\bf D}$
diverges at the origin and so does the energy density \ben T_{00}=
{\cal H} ={\bf E} \cdot {\bf D} -L. \een {\sl Thus this solution
is {\bf not } a smooth soliton solution without sources}. In fact
there is a distributional source \ben {\rm div} {\bf D} = 4 \pi q
\delta ({\bf r}). \een

Finite energy but singular solutions like this of non-linear
theories with distributional sources are a sufficiently
distinct phenomenon
from the familiar finite energy non-singular lump solutions
 without sources as to deserve a different name. The
suggestion has been made \cite{G} that they be called BIons. From the
string point of view the source has a natural interpretation as
being associated with a string ending on a three-brane. In fact
one returns in this way to a picture very close to late nineteenth
century speculations in which an electron is regarded as an
"ether-squirt" on a 3-surface embedded in four dimensional space
\cite{Rouse}. The application to strings is contained in \cite{G} and \cite{CM}.
The present account is largely based on \cite{G}.

\subsection{Maximal Spacelike Hypersurfaces}

Another interpretation of the static solutions
may be obtained as follows.
One introduces the electrostatic potential $\phi=A_0$
and finds the Lagrangian density to be given by
\ben
1-\sqrt{ 1-(\nabla \phi)^2 }.
\een
The Euler-Lagrange equation
\ben
{\rm div} \Bigl (
{ \nabla \phi \over \sqrt
{ 1- (\nabla \phi )^2}  } \Bigr ) =0,
\een
is just that which would be obtained if one sought a maximal
spacelike hypersurface of minkowski spacetime where $\phi$
is now thought of as a time function
\ben
x^0({\bf x}) =\phi({\bf x}).
\een

The maximal hypersurface becomes null at the critical field strength.

\subsection{Catenoids and $D-{\bar D}$ solutions  }

Rather than exciting the electric field we can excite a single scalar $y$.
We get as Lagrangian density
\ben
1-\sqrt{ 1+(\nabla y)^2 }.
\een
The Euler-Lagrange equation
\ben
{\rm div}  \Bigl  ({ \nabla y \over \sqrt{ 1+ (\nabla y )^2}  }
\Bigr ),
\een
is that governing the height function of a minimal surface in four
space-like dimensions.
One readily checks that Monge gauge is not global.
In the spherically symmetric case
There is a branch 2-surface at a finite radius. One needs two Monge patches.
The resulting two sheeted worm-hole or better Einstein-Rosen bridge type
surface looks like two parallel three planes a with finite separation
joined by a neck. The solution is not stable and therefore
one thinks of it as
a Brane-Anti-Brane pair.

\subsection{Charged Catenoids:
$O(1,1)$ symmetry relating Catenoids and Bions} Including both
electric and scalar fields gives a Lagrangian \ben 1-\sqrt{
1+(\nabla y)^2 -(\nabla \phi)^2 }. \een It and the Euler-Lagrange
equations \ben {\rm div} \Bigl ({ \nabla y \over \sqrt{ 1+ (\nabla
y )^2 -(\nabla \phi )^2 }} \Bigr ), \een and \ben {\rm div} \Bigl
({ \nabla \phi \over \sqrt{ 1+ (\nabla y )^2 -(\nabla \phi )^2 }}
\Bigr )=0, \een which are manifestly invariant under an obvious
$O(1,1)$ action analogous to the well-known Harrison
transformation of static Einstein Maxwell theory. Using this
action one may construct everywhere smooth charged catenoids, the
electric field lines passing through the neck or throat in a way
similar to that discussed by Wheeler in the case of
Einstein-Maxwell theory. This family I call under-extreme. They
are obviously analogous to under extreme Reissner-Nordstrom
solutions. One may also excite the scalar field of the BIon
solution. The original flat three-brane acquires a cusp as if it
were being pulled. All of these solutions are singular. I call
them over-extreme. They are obviously analogous to over extreme
Reissner-Nordstrom solutions.

\subsection{The BPS solution: S-Duality}

In the limit of infinite $O(1,1)$ parameter one obtains
an extreme solution analogous to extreme Reissner-Nordstrom.
This solution is in fact supersymmetric. It may be interpreted
as a fundamental (F-) or $(1,0)$ string ending on a three-brane.
Using the electric-magnetic duality one may easily obtain a
magnetic monopole  solution which represents a D-string
or $(0,1)$ string ending on a three-brane.
In fact using $SL(2,{\Bbb Z})$ and the Dirac quantization
condition we can get dyon or $(p,q)$ strings
ending on a three-brane.

\section{Open String Causality}

In string theory, open string states propagating in a background
${\cal F}_{\mu \nu}$ field do so according to a different metric from
the Einstein metric $g_{\mu \nu}$ felt by  closed strong states.

One has
\ben
\Bigl ( { 1 \over g  + {\cal F}  }
 \Bigr ) ^{\mu \nu} =G^{\mu \nu} + \theta ^{\mu \nu},
\een
where $G^{\mu \nu} =G^{(\mu \nu )}$ and $\theta ^{\mu \nu} =\theta ^{[\mu \nu]}$.
If
\ben
G_{\mu \lambda }G^{\lambda \mu }= \delta ^\nu _\mu,
\een
then
\ben
G=g_{\mu \nu} -{\cal F} _{\mu \lambda} g^{\lambda \rho} {\cal F} _{\rho \nu}.
\een

Note that even if
$B_{\mu \nu}=0$ so that ${\cal F}_{\mu \nu} =F_{\mu \nu}$,  the metric $G_{\mu \nu}$ is not invariant under electric-magnetic duality.

\subsection{Boillat metrics}

One may investigate the propagation of
small disturbances of vectors, $A_\mu$ scalars $y$
 and spinors $\psi$
around a Born-Infeld  background using the method of
characteristics. This was done in great detail by Boillat for a
general non-linear  electrodynamic theory. He found that in
general, because of bi-refringence,
 there are a pair of characteristic
surfaces $S={\rm constant} $ satisfying
\ben
\Bigl ( T_{\rm Maxwell}^{\mu \nu} + \mu g^{\mu \nu} \Bigr )
 \partial_ \mu  S \partial _\nu  S=0,
\een where $ T_{\rm Maxwell}^{\mu \nu}$ is the Maxwell stress
tensor constructed form $F_{\mu\nu}$ . Of course the stress tensor
$T^{\mu \nu}$ of the non-linear electrodynamic theory is different
from $ T_{\rm Maxwell}^{\mu \nu}$. The quantity $\mu=\mu (x,y)$
satisfies a quadratic equation whose coefficients depend upon
first and second derivatives of the Lagrangian $L(x,y)$ with
respect to $x$ and $y$. Boillat finds it convenient to fix the
arbitrary conformal rescaling freedom in the characteristic
co-metric by setting \ben C^{\mu \nu}= { 1\over \sqrt{ \mu ^2 -x^2
-y^2} } \Bigl ( \mu g^{\mu} + T_{\rm Maxwell}^{\mu \nu} \Bigr )
\een with inverse or metric \ben C^{-1}_{\mu \nu}=  { 1\over
\sqrt{ \mu ^2 -x^2 -y^2} } \Bigl ( \mu g^{\mu \nu } - T_{\rm
Maxwell}^{\mu \nu} \Bigr ) \een

In general the boundaries of the two  Boillat cones $C_{\rm
Boillat}: C^{-1}_{\mu\nu} v^\mu v^\nu\ge 0, v^0 >0$ and the
Einstein cone $C_{\rm Boillat}: g_{\mu\nu} v^ \mu v^ \nu \ge,
v^0>0$ will touch along the two principle null directions of
$F_{\mu \nu}$ . One sometimes find that one at least of the
Boillat cones lies outside the Einstein cone. In other words small
fluctuations can travel faster than gravitational waves whose
speed is  governed by $g_{\mu \nu}$.

To check causality we examine the Boillat co-cones
$ C^\star_{\rm Boillat} :C^{\mu \nu} p_\mu p_\nu \ge 0, p_0 >0$ and the Einstein
co-cone
$C^\star _{\rm Einstein}: C^{\mu \nu} p_\mu p_\nu  \ge 0, p_0 >0$
in the cotangent space
$T^\star \Sigma_{p+1}$. Suppose that $l_\mu$ is the co-normal
 the Einstein co-cone
$C^\star _{\rm Einstein}$ \ben g^{\mu \nu} l_\mu l_\nu=0. \een The
weak energy condition implies \ben T_{\rm Maxwell}^{\mu \nu} l_\mu
l_\nu \ge 0. \een Thus \ben C^{\mu \nu} l_ \mu l_\nu \ge 0. \een
This means that if $\mu$ is positive then  $C^\star _{\rm
Einstein}$ lies inside or touches $C^\star_{Boillat}$. Remembering
that  duality reveres inclusions one finds then that the Einstein
cone $C _{\rm Einstein}$ lies outside or touches the Boillat cone
$C _{\rm Boillat}$. Note that what we are calling a cone here is
the solid cone. The light cone is the boundary of this solid cone.

\subsection{Hooke's' Law}

Born-Infeld is exceptional in that there is just one solution for
$\mu$:
\ben
\mu =1+x.
\een
Thus there is no bi-refringence. Moreover one finds that
the Boillat co-metric satisfies the remarkable identity
\ben
C_{\rm BI} ^{\mu \nu} = g^{\mu \nu} + T^{\mu \nu}.
\een
I call this identity Hooke's Law for reasons which will be explained below.
Another striking identity is
\ben
\det \bigl ( \delta^\mu_\nu + T^\mu _\nu \bigr )=1.
\een
This follows form another useful identity is
\ben
\det C^{\mu \nu}= \det g^{\mu \nu}.
\een

From Hooke's Law it is easy to see, since $T^{\mu \nu}$ for
Born-Infeld theory satisfies the Weak Energy Condition, that the
Boillat cone lies inside or touches the Einstein cone. In other
words small fluctuations travel with a speed no greater than
gravitational waves. Because the Born-Infeld energy momentum
tensor is invariant under electric-magnetic duality rotations, the
Boillat metric, unlike the open string metric $G_{\mu \nu}$  is
also invariant. One has \ben C^{-1} _{\mu \nu}= { 1\over \sqrt
{1+2x-y^2} } G_{\mu \nu}. \een The conformal factor is related to
the Lagrangian: \ben \sqrt{ - \det ( \eta_{\mu \nu}+ F_{\mu \nu}
)} = \sqrt {1+2x-y^2 }. \een

The reason for the name Hooke's law is that Hooke asserted, in the
days  when the archive was in Latin that {\it ceiiinosssttuu}. In
this way he hoped to make both a priority claim and preserve his
discovery for his own later use.
 Bearing in mind that
{\it u} {\it v} are not distinguished in Latin, the earth
shattering discovery   that he wished  to hide was that {\it ut
tensio sic vis} \cite {J}. In other words stress is proportional to strain.
A standard measure of strain in non-linear elasticity theory is
the difference of two metrics. More precisely, the configuration
space of an elastic medium is a map from an elastic manifold to an
embedding space. There is usually a rest or un-deformed
configuration and one takes as a measure of stress the difference
between the pullbacks from the embedding space to the elastic
manifold in the strained and unstrained configuration.

What we have is an expression involving the difference of two
co-metrics but the idea is similar. One is the co-metric induced
on the brane from the Einstein co-metric and the other is a
measure of the vector field excitations.

\subsection{ Hooke's Law, the Monge-Amp\`ere Equation and Pulse
interactions}

The striking determinantal identity has an interesting application
to the propagation of pulses in Born-Infeld theory.

In flat two dimensional spacetime, the conservation law for the
stress tensor implies that it is given by a single free function,
the  Airy stress function $\psi$, such that \ben T_{tt}=
\psi_{zz}, \quad T_{zz} = \psi_{tt}, \quad T_{tz}=\psi_{zt}. \een
Written in terms of the Airy stress function, the determinantal
identity becomes the Monge-Amp\`ere equation \ben \psi _{zz} \psi
_{tt}- \psi ^2 _{zt}= \psi_{zz}-\psi_{tt} \een This can be solved
exactly (see \cite{pulse} and references therein) by  a Legendre transform under which it becomes
D'Alembert's equation with respect to a new set of variables $T$
and $Z$.

One has \ben T^{tt}={A+B+2AB \over 1-AB}, \een \ben
T^{zz}={A+B-2AB \over 1-AB}, \een \ben T^{zt}={B-A \over 1-AB },
\een where $A=A(T+Z)$ and $B=B(Z-T)$ are arbitrary functions of
their arguments. The relation between the new coordinates $(T,Z)$
and usual coordinates $(t,z)$ is most conveniently expressed using
null coordinates. Let $v+t+z, u=t-z, \xi= Z-T, \eta=Z-T$. The
asymmetrical definition of $\eta$ is so as to agree with
previous work cited in \cite{pulse}, One has \ben dv=d\xi-Bd \eta, \qquad du= -d\eta + A
d\xi. \een Thus \ben (1-AB) d \eta=Adv-du, \qquad (1-AB) d \xi=
dv-Bdu. \een one checks that \ben dT^2-dZ^2= dt^2(1-A-B+AB)
-dz^2(1+ A+B +AB)-2dtdz(A-B) = C^{-1}_{\mu \nu} dx^\mu dx^nu, \een
where \ben C^{\mu \nu}= \eta^{\mu \nu} + T^{\mu \nu}. \een Thus we
see that the Legendre transformation to the new coordinates
$(T,Z)$ used to solve the Monge-Amp\`ere equation in effect passes
to flat inertial coordinates with respect to the Boillat metric.
It should be noted that one does not expect the Boillat metric to
be flat in general.

The general solution consists of two pulses, one right-moving and
one left moving which pass through-another without distortion. In
terms of the usual coordinates $(t,z)$
 they two pulses experience a {\sl delay} That is measured with respect to the closed string metric. However
with respect to the Boillat coordinates, that is measured with
respect to the Boillat metric, there is no delay.

\subsection{Scalars and fermions: Open String Equivalence Principle}

The coupling of scalars has already  been given above. It is easy
to check that the Boillat co-metric determines their fluctuations
around a background, they are in fact governed by the D'Alembert
equation constructed from the co-metric $C^{\mu \nu}$. One may
also consider fermion fields $\psi$. Omitting four-fermion terms,
they couple in a typical Volkov-Akulov fashion.

\ben
- \int dx^{p+1} \sqrt {\det (g_{\mu \nu} + i{\bar \psi} \gamma_\mu \nabla _\nu \psi
+F_{\mu \nu} + B_{\mu \nu} )},
\een
where
\ben
\gamma _\mu \gamma _\nu + \gamma _\nu \gamma_ \mu = 2 g_{\mu \nu}.
\een
Let's define Boillat gamma matrices by
\ben
a^{\mu \nu} = G^{\mu \nu} + \theta ^{\mu \nu},
\een
and
\ben
{\tilde \gamma }^\mu = a^{\mu \nu} \gamma_\nu.
\een

One has
\ben
{\tilde \gamma} ^\mu {\tilde \gamma}^ \nu + {\tilde \gamma }^\nu
{\tilde \gamma} ^ \mu = 2 G^{\mu \nu}.
\een

Because  the leading derivative term in the action is \ben i {\bar
\psi} {\tilde \gamma} ^\mu \partial _\mu  \psi, \een It is clear
that the characteristics of the fermions are also given by the
Open String metric or equivalently the Boillat metric.

Thus we have a sort of
world sheet equivalence principle or universality
holding: all open string fields
have the same characteristics and hence the same maximum speed.

\section{Tolman Redshifting of the Hagedorn temperature}

As an application of the equivalence principle it is interesting
to consider open strings at finite temperature in a background
electromagnetic field. . This was done for the neutral bosonic
string in \cite{FFI}. If the free energy density in the absence of
a background is $F=F(\beta)$ where $\beta$ is the inverse
temperature, then the free energy in a background is obtained by
the replacement \ben F\rightarrow \sqrt{G_{00} }\sqrt{-G_{ij}}
F(\beta \sqrt{G_{00}}), \een where $G_{\mu \nu}$ is the open
strong metric. The first factor may be thought of as a redshift
and volume contraction factor. The rescaling of the argument is
essentially the  Tolman effect whereby in order to retain local
equilibrium in an external static or stationary metric $G_{\mu
\nu}$, the local temperature must vary as $1 \over \sqrt{G_{00}}$.
Note that $G_{00} =1-{\bf E}^2$ and so the redshifting is indeed
{\sl red} shifting and it depends only on the electric field, the
effect diverging at the critical electric field strength.

 Alternatively,
one may regard the effect as being due to the fact that finite
temperature physics corresponds to working in imaginary time with
a period given by the inverse temperature. If the global time
variable is identified with period $\beta$, the  local period will
be
 $\beta \sqrt{G_{00}}$.
Thus the locally measured temperature will be higher. If more than
one metric is involved, then the temperature of states in local
equilibrium may differ, since each will be redshifted by the
appropriate  Tolman factor. In the present case one has closed
string states at temperature $1\over \beta$ and open strong states
at temperature $ 1\over \beta \sqrt{G_{00}}$. The redshifting of
open string states is universal, was confirmed in \cite{T}.

In the absence of a background field the open string has, in
perturbation theory, the free energy has a singularity at the
Hagedorn temperature $T_{\rm Hagedorn}= { 1\over \beta_{\rm
Hagedorn}}$. This is represents a maximum possible temperature
because above it there are so many massive string states that
thermal equilibrium becomes impossible. In a background electric
field the maximum temperature is reduced to \ben T_{\rm Hagedorn}
\sqrt{1-{\bf E}^2 }. \een

This effect has been interpreted as being due to a reduction in
the effective string tension in an electric field. This is
certainly true but one cannot derive the exact formulae from that
assumption alone whereas everything follows rather naturally by an
application of the equivalence principle, as long as one uses the
open string metric.

\subsection{Shocks and Exceptionality}

Loosely speaking, shocks can occur of the speed  of waves depends
on the phase or amplitude in such a way that different waves
surfaces $S={\rm comstant}$ can catch up and form caustics. More
precisely one assumes the ansatz \ben F_{\mu \nu} = { F^0}  _{\mu
\nu}  \bigl ( f(S) \bigr ) \een with \ben S= {\bf n} \cdot {\bf x}
- v({\bf n}, S)t. \een and $f(S)$ an arbitrary function. The
surfaces $S={\rm constant}$ are hyperplanes and are to be thought
of as surfaces of constant phase. If the phase speed $v$ depends
non-trivially on the phase $S$ there will be shocks along the
envelope of the hyperplanes. Theories without shocks for which $v$
is independent of $S$ are called exceptional.

Boillat has shown that the only form of non-linear electrodynamics
with a sensible weak field limit is that of Born-Infeld.

Theories with shocks  are essentially incomplete. In a sense, like
General Relativity they predict their own demise. By contrast
Born-Infeld, like classical Non-Abelian Yang-Mills theory seems to
be a perfect example of a classical theory. As far as one can tell
it appears to possess the property which is known to be true for
Yang-Mills theory, that regular Born-Infeld  initial data with
finite energy may evolved for all time to give everywhere
non-singular solutions of the field equations. For a more detailed
discussion and references to the original literature see \cite{GH}

\section{Strong Coupling Behaviour of Born-Infeld}

There are (at least) two interesting strong coupling limits
of Born-Infeld theory.
\medskip \begin{itemize}
\item A Weyl-invariant duality  invariant theory
which appears to be related to a fluid of massless magnetic Schild
type strings and may describe string theory near critical electric
field strengths.

\item A massive theory which  is related to a
fluid of massive strings and may be related to current ideas about
$D-{\bar D}$ annihilation and tachyon condensates.

\end{itemize}
In both cases the key to understanding these  limiting theories is
passing to the Hamiltonian formulation. It also helps to bear in
mind some facts about:

\subsection{Simple 2-forms, Distributions and String Fluids}

A 2-form $\Omega$ is simple iff \ben \Omega =\alpha \wedge \beta
,\een
 equivalently
\ben \Omega \wedge \Omega=0.\een In particular since the matrix of
components has $\Omega _{\mu \nu} $ has rank two: \ben \det \Omega
_{\mu \nu} =0.\een In four spacetime dimensions $\Omega$ is simple
iff \ben \Omega _{\mu \nu} \star \Omega ^{\mu \nu}=0. \een Of
course $\alpha$ and $\beta$ are not unique but a field of simple
2-forms defines the  unique two-dimensional sub-space which they
span in the cotangent space $T_xM$ at every point of spacetime.
Hence, given a metric, a simple two form is equivalent to a simple
bi-vector $\Omega_{\mu \nu}$ which  defines a distribution $D$ of
2-planes in the tangent space $TM$. Raising indices with the
metric, the simplicity condition becomes in terms of the bi-vector
\ben \Omega ^{[\mu \nu} \Omega ^{\alpha ]\beta}=0.\een
 One
may think of the distribution $D$ as a sub-bundle of the tangent
bundle with two-dimensional fibres. The 2-planes will be timelike,
null or spacelike depending upon whether 
$\Omega_{\mu\nu} \Omega ^{\mu \nu}$ is negative, zero
or positive respectively. (Note that this statement 
is signature indepbedent.)  In the timelike case one may chose
$\alpha$ to be timelike and $\beta$ to be spacelike. In the null
case one may choose $\alpha$ to be null and $\beta$ to be
spacelike.

In general the distribution $D$ will not be integrable. That is
neighbouring 2-planes will not mesh together to form  the tangent
spaces of a co-dimension two family of   2-dimensional surfaces.
If it is, then  if two vector fields $X$ and  $Y$ belong to $D$
then their Lie bracket $[X,Y]$ must belong to $D$. Such an
integrable distribution may be identified as a gas or soup,
perhaps more accurately a spaghetti of strings. The condition for
integrability may be expressed in various ways. For us the
simplest condition is in terms of the bi-vector and is
 \ben
\Omega ^{[\alpha \beta}
\partial _\kappa \Omega ^ {\mu ]\kappa}=0. \label{integr}\een
 Note that if $f$ is a smooth function, then $\Omega$ and
$f\Omega$ define the same distribution and if the first
 is
integrable then so is the second. Moreover, the partial derivative
in (\ref{integr}) may be replaced by a torsion free covariant
derivative. In four spacetime dimensions we may re-express the
integrability condition as
 \ben \star \Omega_{\mu \nu} \nabla
_\kappa \Omega ^{\nu \kappa}=0. \een

We may re-write this as
 \ben \Omega \wedge \delta \Omega=0.\een

where $\delta \Omega = \star d \star \Omega$.

Now if we take for $\Omega$ the Amp\`ere tensor $K_{\mu \nu}$ of
any non-linear electrodynamic theory. We see that any simple
solution of the equations of motion \ben \nabla K^{\mu \nu}=0\een
automatically defines an integrable distribution. In other words
non-linear electrodynamic theory  supplemented with the constraint
\ben F \wedge K=0, \een may be re-interpreted as a (vorticity
free) string fluid. Different Lagrangians correspond to different
equations of state.

\subsection{0-brane fluids} This section is based on part on \cite{pulse}
 The situation described  above  should
be compared with  the familiar case of a non-linear scalar field
theory with a Lagrangian $L(\partial \phi)$ containing no explicit
dependence on the scalar field $\phi$. The equations of motion may
be cast in the form

\ben \nabla_\mu  (s U^\mu ) =0 \een
 where $U^\mu$ is a normalized timelike vector  given by
 \ben
U_\mu ={\partial_\mu \phi  \over \sqrt { (\partial \phi )^2 } }.
 \een
and \ben s U^ \mu = { \partial L \over \partial (\partial _\mu
\phi ) } \een may be interpreted as a conserved entropy current.
The quantity $s$ corresponds to the entropy density and $\rho$ to
the local energy density. The energy momentum tensor takes the
perfect fluid form \ben T^{\mu \nu}= (\rho +P ) U^\mu U^\nu -P
g^{\mu \nu} \een One has \ben P=L .\een If one defines \ben T^2
=(\partial \phi) ^2, \een it is natural to regard the pressure as
a function of the temperature $T$  but the energy density as a
function of the entropy density $s$. In fact they are related by a
Legendre transform.  One finds that \ben \rho + P=sT \een and \ben
s = {\partial P \over
\partial T} \qquad T = {\partial  \rho \over \partial s} \een

It is an illuminating exercise to convince oneself that finding
the speed of small fluctuations by the calculating the sound speed
\ben c_s= \sqrt {\partial P \over
\partial \rho}\een is equivalent to calculating the characteristics
, that is the Boillat metric.

The most interesting case from the present point of view arises
when one takes the scalar Born-Infeld Lagrangian \ben L=1- \sqrt{
1- (\partial \phi)^2 }. \een

One has \ben P= 1-\sqrt{1-T^2}\een and \ben \rho ={ 1 \over
\sqrt{1+ s^2} }-1 \een which has a maximum temperature reminiscent
of the Hagedorn temperature. However the detailed equation of
state is different. One has the equation of state \ben P= {\rho
\over 1 + \rho} \een and hence \ben c_s= {1 \over 1+\rho}.\een

Note that one need not regard the conserved current as an entropy
current if one does not wish to. One could regard it as a
conserved particle number.

\subsection{The Weyl-invariant Bialynicki-Birula limit}

The Hamiltonian density , with units restored is

\ben
{\cal H} = T^2 \sqrt{ 1+ {{\bf B}^2 + {\bf D}^2 \over T^2}
 +{ ({\bf D}\times {\bf B})^2 \over T^4 } } -T^2.
\een
One can take the limit $T\downarrow 0$ to get
\ben
{\cal H} = |{\bf D} \times{\bf B} |. \label{hammy}
\een
This gives the constitutive relations
\ben
{\bf E}= -{\bf n} \times {\bf B}, \qquad {\bf H}= {\bf n} \times {\bf D},
\een
where we have defined a unit vector in the direction of the Poynting vector
$ {\bf D} \times{\bf B} $
\ben {\bf n}= {{\bf D} \times {\bf B} \over  |{\bf D} \times{\bf B} | }. \een
Remarkably these constitutive relations (which arise as the limiting form of the constitutive
relations of the full theory) imply the constraints
\ben
{\bf E} ^2 -{\bf B}^2 =0 \qquad {\bf E}.{\bf B} =0. \label{conny}.
\een

Defining a  null vector $l^\mu=(1, {\bf n}) $, the energy momentum tensor becomes
\ben
T^{\mu \nu} = {\cal H} l^\mu l^\nu.
\een
It follows that the trace vanishes
\ben
T^\mu _\mu=0,
\een
and hence the limiting theory is Weyl-invariant.
 It may be checked that is Lorentz-invariant and invariant under
electric-magnetic duality rotations. One may also check from the
equation of motion that there are infinitely many conserved
symmetric  tensors \ben T^{\mu_1 \mu_2 \dots \mu_k}={\cal H}
l^{\mu_1} l^{\mu_2} \dots l^{\mu_k}. \een

The constraints (\ref{conny}) tell us that the Faraday
tensor $F_{\mu \nu}$ is simple
\ben
\det F_{\mu \nu}=0,
\een
and null,
\ben
F_{\mu \nu} F^{\mu \nu}=0.
\een
Thus $F_{\mu \nu}$  defines a two plane
which is null, that is,  the two-plane is  tangent to the light
cone along the lightlike  vector $l^\mu$ and
\ben
F_{\mu \nu} l^\mu=0.
\een
The equations of motion tell us that the two-plane distribution
in the tangent space  defined by the Faraday two-form $F_{\mu \nu}$
is integrable, that is surface forming, and hence that spacetime is foliated by two-dimensional
lightlike surfaces which may be interpreted as the world sheets
 of magnetic null or Schild strings.  In other words, in this critical limit
which may be interpreted as describing Born-Infeld theory near
critical field strength, the system dissolves into a gas or fluid
of Schild strings.

Since electric-magnetic duality is maintained in the limit, one can of course pass to a dual description
in terms of $K_{\mu \nu}$.
This amounts to the observation that ${\bf H}^2 ={\bf D} ^2$ and ${\bf H }.{\bf D}=0$, i.e. $K_{\mu \nu}K^{\mu \mu}=0$
and $K_{mu \nu } \star K^{\mu \nu} =0$.

\subsection{Covariant formulation of UBI using auxiliary fields}

The Weyl-invariant limit was called by Bialynicki-Birula
\cite{B1,B2}, Ultra-Born-Infeld. Let us follow him and consider
\ben L= -{\mu \over 4} F_ {\mu \nu}
 F^{\mu \nu}  + { \nu \over 4} F_{\mu \nu} \star F^{\mu \nu},
\een where $\mu$ and $\nu$ are dimensionless auxiliary fields,
variation with respect to which gives the constraints \ben F_{\mu
\nu} F^{\mu \nu}=0=F_{\mu \nu} \star F^{\mu \nu}. \een Variation
with respect to $A_\mu$ gives the field equation.

Note that in axion-dilaton Maxwell theory, the auxiliary fields
could be functions of the dimensionless dilaton $\Phi$ and axion
$\chi$, $\mu=\mu(\Phi, \chi)$, $\nu(\Phi,\chi)$ chosen in such a
way that the system was $SL(2,{\Bbb R})$ invariant. The dilaton
and axion  provide
 a  map from  spacetime into
$SL(2,{\Bbb R} )/SO(2)$ and  one would, in general,  have a
non-linear sigma model type  kinetic term for them (see e.g.
\cite{GR2}). For dimensional reasons it must be multiplied by $T$.
 In the limit we are considering the kinetic term vanishes and the
axion and dilaton become auxiliary fields.

\subsection{Tachyon Condensation}

This subsection is based on \cite{GHY}where references to the
string literature may be found. The basic idea goes back to Ashoke
Sen. In the presence of a tachyon field the Born-Infeld Lagrangian
density is believed to be modified by the tachyon potential $V$ to
take the form \ben L= V-V \sqrt {1-{\bf E} ^2 + {\bf B} ^2 - ({\bf
E} \cdot {\bf B})^2 }. \een We are now keeping $\alpha^\prime$
 fixed and using units in which $ 2 \pi \alpha ^\prime=1$. Now it is believed that $V$ has a critical
point away from zero
a which $ V$ vanishes. It is also believed that dynamically the
system will relax to the state with $V=0$, a so-called tachyon condensate.
 One may thus ask, what happens
 to the Born-Infeld  vector in this limit. Again the Lagrangian density
causes confusion: it vanishes identically in the limit. However
the Hamiltonian density is \ben {\cal H} = \sqrt{ V^2(1 + {\bf
B}^2) + {\bf D}^2
 +  ({\bf D}\times {\bf B})^2  } - V.
\een

and the limiting form is
\ben
{\cal H} = \sqrt{  {\bf D}^2
 + ({\bf D}\times {\bf B})^2  }.
\een

The resulting constitutive relations are
 \ben
{\bf H} = { {\bf B} {\bf D}^2 -{\bf D} ( {\bf B}.{\bf D}) \over
\sqrt{ {\bf D^2} + ({\bf B} \times {\bf D} )^2 }} \een

 \ben
{\bf E} = { {\bf D} +{\bf D} {\bf B}^2 -{\bf B} ( {\bf B}.{\bf D})
\over \sqrt{ {\bf D^2} + ({\bf B} \times {\bf D} )^2 }} \een

 They tell us that ${\bf E} \times {\bf H}= {\bf D} \times {\bf B}$,
 and therefore
 the theory is Lorentz-invariant. One may check that electric-magnetic duality invariance is lost in this limit.
  The constitutive relations  also imply
that \ben {\bf D}. {\bf H}=0. \een but \ben {\bf D}^2 -{\bf H}^2
>0. \een

It follows that the Amp\`ere tensor $K_{\mu \nu}$ with components
${\bf D}, {\bf H}$, is simple but timelike. Thus the two-form $
K_{\mu \nu}$ it defines a 2-plane distribution in the tangent
space. As discussed above the equation of motion for $K$ implies
that the distribution is integrable.

The limiting theory maybe expressed in terms of the Amp\`ere
tensor $K_{\mu \nu}$. One  way to proceed is to consider a dual
Lagrangian. We define $G=\star K$. The field equation $d\star K=0$
becomes the Bianchi-Identity $dG=0$. We now set $G=dC$ and
consider the Lagrangian \ben {\hat L}= \sqrt{ { 1\over 2} G_{\mu
\nu} G^{\mu \nu } }= \sqrt{ - { 1\over 2} K_{\mu \nu} K^{\mu
\nu}}. \een

The action is now varied with respect to $ C$ but {\sl subject to
the constraint}  that \ben K_{\mu \nu} \star K^ {\mu \nu }=0. \een

The resulting energy momentum tensor is given by \ben T_{\mu\nu} =
-{ K_{\mu \lambda} K_\nu \thinspace ^\lambda \over \sqrt{ -{ 1
\over 2} K_{\sigma \tau } K^{\sigma \tau} }} \een

The trace is given by \ben T^\mu _\mu= -\sqrt{ -2 K_{\sigma \tau}
K^{\sigma \tau} }. \een and therefore  this is certainly not a
conformally invariant theory.

Locally one may pass to a rest frame in which ${\bf B}=0$. Then
\ben {\cal H} = |{\bf D} |. \een This is precisely what one
expects of electric flux tubes with an energy proportional to the
length and to the total flux carried by the tube.

In this rest frame one finds that
 \ben
T_{\mu \nu} = \pmatrix { \tau & 0 & 0& 0 \cr 0 & -\tau & 0 &0 
\cr 0 &0 &0 & 0   \cr 0 &0  &0 & 0 \cr }.
 \een

This is just what one expects for a string fluid.

\section{The M5-brane}

In this concluding section I will indicate how many of the ideas
described above extend to the theory of the M5-brane. To
paraphrase Hooke {\it ut D3-brane sic M5-brane}. Indeed from the
M-Theory point of view one should perhaps have reversed the logic,
since one may regard the equations of Born-Infeld theory as the
dimensional reduction of the M5-brane equations. The theory and
it's equation have a reputation for complexity and so I will try
to present them in as direct a way as possible. The interested
reader may find references to the original papers and the
statements made below the paper on which section is based
\cite{GibWest}.

One is of course
considering a 6-dimensional non-linear theory involving scalar and spinor
 fields and in addition and closed 3-form $H_{\alpha \beta \gamma}$. In what follows
I shall follow the original papers except that  $\mu =0,1,\dots
,5$. In particular in this section I  shall follow their lead  in
this section be using the mainly positive signature
 convention.
\subsection{Bianchi Identity}
Consider the simplest situation: just the 3-form in a fixed
background Einstein-metric $g_{\mu \nu}$. Thus \ben dH=0. \een
Locally therefore  one has $H=dA$, for some 2-form $A$.
\subsection{Non-Linear self-duality}
Now in six-dimensional Minkowski spacetime and acting on
three-forms the standard linear Hodge duality is an involution of
order two:$\star \star =1$ and in linear theory self-duality is a
consistent field equation. In other words closure of $H$ and the
self-duality condition give the {\sl complete} set of equations of
motion. For the M5-brane a very remarkable {\sl non-linear
self-duality}
 condition  is possible which fulfills the same purpose.
This was first discovered by Perry and Schwarz and its covariant form
written down by Howe Sezgin and West.
One does not seem to be able to construct a covariant Lagrangian just
using the 2-form $A$. Non-covariant variational principles
exist and a
covariant action principles has been written down
using an additional scalar field which acts as a time function.
For the time being we need in these lectures only the equations of motion.

This remarkable condition is perhaps most expeditiously written as
\ben \star H_{\alpha \beta \gamma} = { 1\over \sqrt {1 + {2 \over
3} H^2}} \Bigl [ (1+ {4\over 3} H^2 ) \delta ^\epsilon_\alpha-4
(H^2) ^\epsilon _\alpha\Bigr] H_{\epsilon \beta \gamma}. \een Of
course in the limit of small $H$ we have $H\approx \star H$. As
stated above, if one reduces to five spacetime dimensions the
equations reduce to the standard Born-Infeld equations.

\subsection{Boillat Cone and Hooke's Law}
One may introduce the analogue of the Boillat co-metric: \ben
C^{\alpha \gamma}= { Q \over (2-Q) } \Bigl ( g^{\alpha \gamma} (1+
{4 \over 3} H^2) -4 (H^2) ^{\alpha \gamma}  \Bigr ), \een where
\ben Q=-{ 3 \over H^2} ( 1+ { 2 \over 3} H^2 ). \een

The characteristics of the scalar, spinor and 3-form equations of
motion are determined by the Boillat metric $C^{\mu \nu}$. Moreover one
may introduce an energy momentum tensor $T^{\mu \nu }$ which satisfies
Hooke's Law: \ben T^{\alpha \beta }=g^{\alpha \beta } -C^{\alpha
\beta }, \een and is conserved \ben T^{\mu \nu }\\_{;\nu}=0. \een
Note the sign change in Hooke's law because of the signature
change. (However  $T^{00} \ge 0$ in both conventions.)

One may prove that $T^{\mu \nu}$ satisfies the Dominant Energy
Condition and hence, as with Born-Infeld theory, that the Einstein
cone never lies inside the Boillat cone. In general the two cones
touch along a circle of directions.

\subsection{Weyl-invariant strong coupling limit}

In general the trace of the energy momentum tensor $T^\mu _\mu$
does not vanish. The theory is not Weyl-invariant except at
vanishing field strength. However it becomes Weyl invariant in the
limit of strong coupling.
 As with Born-Infeld, the most direct
route to this result is the non-covariant (in our case  $SO(5)
\subset SO(5,1)$ symmetric) form of the equations. One defines a
pair of two-forms $E_{ij}= H_{0ij}$ and $B_{ij}=-{ 1\over 6}
\epsilon_{ijpqr} H^{pqr}$. The Bianchi identity may be written in
an obvious notation as
 \ben {\partial {\bf B}  \over \partial t
} + {\rm curl } {\bf E} =0, \qquad {\rm div} {\bf B}=0. \een To
close the system one need a constitutive relation. To this end one
defines \ben {\cal H}= \sqrt{ \det (\delta_{ij} + 4 B_{ij} ) }-1.
\een The full non-linear self duality constraint has as solution
\ben E_{ij}= { 1\over 16} {\partial {\cal H} \over
\partial B_{ij}}.
 \een The quantity ${\cal H}$ is the energy density $T_{00}$.

One may now  restore dimensions by setting \ben {\cal H}= T^2
\sqrt{ \det (\delta_{ij} + 4 {B_{ij} \over T} ) }-T^2. \een where
$T$ has dimension mass cubed. The limit $T \downarrow$ is now
easily taken. More interesting than the general formulae  for
$E_{ij}$ are the results for the energy momentum tensor. It takes
the null matter form \ben T^{\mu
 \nu} = {\cal H} l^\mu l^\nu, \een
where, $l^\mu$ is again a null vector in the direction of the
Poynting flux. Thus, just as is the case with Born-Infeld in fours
spacetime dimensions, we attain Weyl-invariance in this limit and
the theory has infinitely many conservation laws.

Point wise, one may skew diagonalize $B_{ij}$. In general it has
rank four and two distinct skew eigenvalues $B_1$ and $B_2$
respectively. Of course the basis in which $B_{ij}$ is skew
diagonalized will in general vary with position. Pointwise one
finds that if $l^\mu=(1,0,0,0,0,1)$

\ben H= (dt-dx^5) \wedge(B_2 dx^2 \wedge dx^2 + B_1 dx^4 \wedge
dx^5). \een

The three form $H$ is in general not self-dual and is the sum of
two totally simple three-forms. One factor is the null one form
$L_\mu dx^\mu$, and one has   $H_{\mu \nu \sigma} l^\mu=0$, $
\star H_{\mu \nu \sigma} l^\mu=0.$

The quantum mechanical  nature of this 
mysterious conformally invariant theory is an interesting 
challenge for the future.


\begin{thebibliography} {99}
\bibitem{DB} L de Broglie, Un Itin\'eraire Scientifique, {\it \'Editions
Scientifique} (1987) 
\bibitem{GGHT} JP Gauntlett, G W Gibbons, C M Hull and P Townsend,
BPS Staes of D=4 N=1 Supersymmetry {\it Comm Math Phys} {\bf 216} (2001) 431-459

\bibitem{GWG} G W Gibbons, Convex Cones in Physics, unpublished notes
of lectures delivered  at the Yuakawa Insitute

\bibitem{H} S W Hawking, The conservation of matter in general
relativity {\it Comm Math Phys} {\bf 18} (1970) 303-306 
\bibitem{HE} S W Hawking and GFR Ellis, {\it The Large Scale Structure of Spacetime} (1973) , Cambridge University Press 
\bibitem{Z1} Ya B Zeldovich, The Creation of Particles and 
Anti-Particles in Electric and Gravitational Fields {\it Magic Without Magic}, ed J R Klauder
(277-288)

\bibitem{Z2} Ya B Zeldovich and L P Pitaevsky, On the possibility of the 
Creation of Particles by a Classical Gravitional Field {\it Comm Math Phys} {\bf 23} (1971) 185

\bibitem{Monge} G Monge, Application de l'analyse  a\` la geom\'etrie (1807)
\bibitem{Rouse} W Rouse Ball, A Hypothesis relating to the Nature
of the Ether \& Gravity {\it Messenger of Mathematics} {\bf 21}
(1891) 20-24

\bibitem{J} L Jardine {\it Ingenious Pursuits} Little, Brown and Company  
(1999)

\bibitem{B1} I Bialynicki-Birula, Non-linear Electrodynamics:
Variations on a Theme of Born and Infeld, in {\it Quantum Theory
of Fields and Particles} eds B. Jancerwicz and J Lukierski, World
Scientific, Singapore (1983)
\bibitem{B2} I Bialynicki-Birula, Field Theory of a Photon Dust,  {\it Acta Physica Polonica} {\bf B 23} (1992) 553-559
\bibitem{FFI} E J Ferrer, E S Fradkin and V de la Incarra, Effect
of a background electric field on the Hagedorn temperature {\it
Phys Lett }{\bf B 248} (1990) 281-287

\bibitem{T} A A Tseytlin, {\it Nucl Phys} {\bf B 460} (1998)  69 
\bibitem{G} G W Gibbons {\it Nucl Phys} Born-Infeld particles nad
Dirichlet p-branes  {\bf 514} (1998) 603-639 {\tt 9709027}
\bibitem{pulse} G W Gibbons, Pulse Propagation in Born-Infeld Theory {\tt
hep-th/0104015}
\bibitem{GibWest} G W Gibbons and P C West, The metric and strong
coupling limit of the M5-brane  {\it J Math Phys} {\tt hep-th/0011149}
\bibitem{GH} G W Gibbons and C A R Herdeiro, {\it Phys rev} {\bf D 63} (2001) 064006-1 - 064006-17 {\tt hep-th/0008052}


\bibitem{GR1} G W Gibbons and D A Rasheed {\it Nucl Phys} {\bf B454} (1995) 185-206 {\tt hep-th/}



\bibitem{GR2} G W Gibbons and D A Rashhed {\it Phys Letts} {\bf B 365} (1996) 46-50 {\tt hep-th/ }


\bibitem{GHY} G W Gibbons K Hori and P Yi, String Fluid from Unstable D-brane, {\tt hep-th/0007019}


\bibitem{CM} C G Callan and J Maldacena, Brane
Dynamics from  the Born-Infeld action, {\it Nucl Phys } {\bf B 513}
(1998)
 198 {\tt hep-th/9708147} 


\bibitem{GK} G W Gibbons and K Hashimoto, 
Nonlinear Electrodynamics in Curved Backgrounds {\tt hep-th/0007019}



\end{thebibliography}
\end{document}